\documentstyle[preprint,aps,pre,epsf]{revtex}

\newcommand{\myfig}[3] {
\begin{figure}
\centerline{
\epsfysize=#2  \epsfbox{#1.eps}
}
\caption{#3}
\label{#1}
\end{figure} }

\def\bfx {{\bf x}}
\def\bfk {{\bf k}}
\def\bfp {{\bf p}}

\def\baral { \bar{\alpha}}
\def\barbe { \bar{\beta}}
\def\barga { \bar{\gamma}}
\def\barpa { \partial}

\def\comm {,\,\,\,\,\,}
\def\calh { {\cal H}}

\def\be{ \begin{equation} }
\def\ee{ \end{equation} }

\begin{document}
\tightenlines
\draft
\title{\bf
Monte Carlo Renormalization Group Analysis of Lattice
$\phi^4$ Model in $D=3,4$}
\author{M.~Itakura}

\address{Center for Promotion of Computational Science and Engineering,
Japan Atomic Energy Research Institute,
Meguro-ku, Nakameguro 2-2-54, Tokyo 153, Japan\\}
\date{\today}
\maketitle

\begin{abstract}
We present a simple, sophisticated method to capture 
renormalization group flow  in Monte Carlo simulation,
which provides 
important information
of critical phenomena.
We applied the method to  the
$D=3,4$ lattice $\phi^4$ model
and obtained a renormalization flow diagram
that well reproduces theoretically predicted  behavior of 
the continuum $\phi^4$ model.
We also show that the method can be easily applied to 
much more complicated models, such as frustrated spin models.
\end{abstract}
\pacs{PACS numbers:  02.70.Lq, 75.10.Hk}

\section{Introduction}
Renormalization-group (RG) theory \cite{kadanoff,wilson,amit,phi6}
has drastically improved our perspective
about phase transition. 
When it is combined with Monte Carlo (MC) simulations,
it becomes a very powerful tool to investigate critical phenomena.
However, the so-called Monte Carlo renormalization group (MCRG)
method 
\cite{mcrg,mcrg1,mcrg2,mcrg3,fftrg}
requires an elaborate scheme and experience,
and  application has been restricted to simple models
such as the Ising ferromagnet up to now.
In this paper 
we reformulate the MCRG method and present 
a simple way to obtain the RG flow diagram in a MC simulation,
which can provide essential information about critical phenomena.
This paper is organized as follows.
In Sec. II, problems in the conventional MCRG scheme are
explained and remedies for each problem are presented.
In fact, modification of the conventional scheme
leads to use of Binder's parameter \cite{binder} and
the second parameter, which is used in 
recent high-precision  numerical analyses \cite{hasenbusch,ballesteros}.
In Sec. III,
expected behavior of RG flow in the $D=4$ lattice  $\phi^4$
model is presented for  comparison with the MC result.
Some caution on MC simulation just at the upper critical
dimension is also presented.
In Sec. IV, details of the MC simulation
of the lattice $\phi^4$ model are described.
In Sec. V and VI, results of the MC simulation for $D=3,4$
are presented.
In Sec. VII,
we summarize the result and 
discuss 
possible applications of the method
to more complicated models.

\section{ Modification of the conventional MCRG method}
In the conventional MCRG scheme \cite{mcrg,mcrg1,mcrg2,mcrg3},
critical exponents are calculated from eigenvalues of the 
linearized RG transformation:
\begin{equation}
{
\partial K_i(L,b_1)
\over
\partial K_j(L,b_2)
\label{rgm1}
}
\end{equation}
where $K_i(L,b)$ denotes coupling strength of the $i$'th term of
a block-spin Hamiltonian with block size $b$, and $L$ denotes
the original size of the system.

One problem is that one should use the proper definition of block spin,
otherwise $K_i(L,b)$ goes to zero or infinity as $b$ becomes large 
even at the critical point.
For the Ising model, the majority rule is one of the options.
However, it cannot be extended to more complicated models.
There are infinite kinds of definitions and
location of the fixed point 
(along an irrelevant direction) depends on it \cite{mcrg1}.

In this paper, we use one that seems most simple and 
suitable  for Monte Carlo simulation,
\be
\bar{s_b} \equiv s_b/\sqrt{<s_b^2>}, \label{sbar}
\ee
where $s_b$ denotes summation of all spins in the block.
This definition is implicitly used in the definition of Binder's
parameter.
Actually, we use coarse-grained spin obtained by 
cutting high-momentum modes off, as in Ref.\cite{fftrg}.
However, we set $b=L$ at the last stage, therefore
real-space RG and momentum-space RG do not make
much difference.

Another problem is that the 
block size $b$ should be much smaller than $L$, otherwise
the behavior of the block spin is affected by 
a boundary condition that is nonuniversal. 
Thus one faces a tradeoff that for larger $b$, 
behavior of block spins become more asymptotic 
(couplings of irrelevant terms become smaller)
but their behavior deviates from bulk one.
The solution is to use the following matrix 
\be
{\partial K_i(bL,bL)
\over
\partial K_j(L,L)
}
=
{
\partial K_i(b L,b L)
\over
\partial K_j(b L,b)
}
\cdot
{
\partial K_i(b L,b )
\over
\partial K_j(b L,1)
}
\cdot
{
\left[
\partial K_i(L,L)
\over
\partial K_j(L,1)
\right]^{-1}
}
,
\label{rgm2}
\ee
instead of (\ref{rgm1}).
At the  critical point,
\be
{
\partial K_i(b L,b L)
\over
\partial K_j(b L,b)
}
=
{
\partial K_i(L,L)
\over
\partial K_j(L,1)
}
\ee
should be satisfied and
eigenvalues of the matrix (\ref{rgm2})
coincide with that of (\ref{rgm1}).
Thus one can use as large a block size as the system size.

Yet another, and most severe problem, 
is referred to as the "redundancy problem" \cite{mcrg2,mcrg3}.
In the conventional MCRG scheme,
one should observe very large numbers of block-spin interaction terms
to construct  a fixed-point Hamiltonian.
However, in the continuum $\phi^4$ theory, there are only two kinds of
independent coupling constants (mass and coupling).
The lattice Hamiltonian approaches this continuum asymptotically
after the momentum-space RG transformation, and
the two terms are enough to observe essential RG flow.

Let us consider  the Hamiltonian defined on 
$D$-dimensional continuum space:
\be
\calh=\int d\bfx \left[ {\gamma\over 2} \left[\nabla \phi(\bfx)\right]^2+
 {\alpha \over 2} \phi(\bfx) ^2 +\beta \phi(\bfx)^4 \right].
\label{hcont}
\ee
Note that it is  not a 
microscopic Hamiltonian: it is a phenomenological 
Hamiltonian and
the parameters $\gamma,\alpha,\beta$ should be
determined to reproduce experimental results.
In other words, it is the logarithm of probability
of observing a specific configuration of some physical quantity
$\phi(\bfx)$ in the experiment.
This means that short length scale fluctuations 
beneath the resolution $l$ of the observation device
are already integrated out and absorbed into the 
parameters $\gamma,\alpha,\beta$.
Thus parameters depend on the cutoff length $l$
 and should be denoted by
 $\gamma_l,\alpha_l,\beta_l$.
We also denote the physical quantity $\phi(\bfx)$ averaged 
over a volume of linear length $l$ by $\phi_l (\bfx)$.

We replace the parameters in (\ref{hcont}) by
"regularized" ones defined as follows:

$$\bar{\alpha}_l= l^D <\phi_l(x)^2> \alpha_l \comm
\bar{\beta}_l=l^D <\phi_l(x)^2>^2 \beta_l \comm
\bar{\gamma}_l=l^{D-2} <\phi_l(x)^2> \gamma_l.$$
In terms of this regularization, the Hamiltonian is expressed as
\begin{eqnarray}
\int _{|\bfp|\leq \pi} d\bfp 
{\baral+\barga \bfp^2 \over 2}
\bar{\phi}(\bfp)\bar{\phi}(-\bfp)
\nonumber \\
+\barbe \int _{|\bfp_i|<\pi} d\bfp_1 d\bfp_2 d\bfp_3 d\bfp_4
\delta(\sum_i \bfp_i )
\bar{\phi}(\bfp_1)\bar{\phi}(\bfp_2)
\bar{\phi}(\bfp_3)\bar{\phi}(\bfp_4),
\label{hcrf}
\end{eqnarray}
where $\bar{\phi}(\bfp)\equiv \phi(\bfp l)<\phi_l^2 >^{-1/2}$.
All possible values  of the regularized parameter 
fall on a two-dimensional manifold
on which $<\bar{\phi}_l(x)^2>=1$ is satisfied.
This regularization is suitable for MC simulations, compared to
the  field theoretical one, $\gamma \equiv 1$.
Figure \ref{rgf3} and Fig.\ref{rgf4} show
schematic RG flow of these regularized parameters
in $D=3$ and $4$, respectively:
high temperature, low temperature, Gaussian, and Wilson-Fisher fixed point 
are denoted by H,L,G, and WF, respectively.

Now let us return to the lattice model defined as follows:
\be
H_L = {\gamma \over 2}\sum_{<ij>} (s_i-s_j)^2 +\sum_i 
{\alpha\over 2}s_i^2 +\beta s_i^4.
\label{hlat}
\ee
where the summation $\sum_{<ij>}$ suns over all nearest neighbor pairs.
If we apply the momentum-space RG transform of factor $b$ to (\ref{hlat}),
the renormalized Hamiltonian takes the following form:
\begin{eqnarray}
H(L,b)&=&
{\baral(L,b) \over 2} (L/b)^{-D} \sum_{|\bfp |\leq \pi}^{(L/b)}
 \bar{s_b}(\bfp)\bar{s_b}(-\bfp) \label{hlba} \\
&+& {\barga(L,b) \over 2} (L/b)^{-D} \sum_{|\bfp |\leq \pi}^{(L/b)}
 \bfp^2 \bar{s_b}(\bfp)\bar{s_b}(-\bfp) \label{hlbc} \\
&+&\barbe(L,b) 
(L/b)^{-3D} \sum
^{(L/b)}_{
\begin{array}{c}
{\scriptstyle |\bfp_i| \leq \pi }\\
{\scriptstyle \bfp_1+\bfp_2+\bfp_3+\bfp_4=0 }
\end{array}
}
 \bar{s_b}(\bfp_1)\bar{s_b}(\bfp_2)\bar{s_b}(\bfp_3)\bar{s_b}(\bfp_4),
\label{hlbb}
\end{eqnarray}
where $\bar{s}_b(\bfp)=s(b \bfp) <s_b^2(x)>^{-1/2}$ and
the symbol $\sum_\bfp^{(N)}$ denotes summation over
$ 0,\pm 2\pi/N,\pm 4\pi/N,\cdots,$ for each component of $\bfp$,
$s(\bfp)$ denotes Fourier components of $s_i$, and 
\be
s_b(\bfx)= (L/b)^{-D} \sum _{|\bfp|\leq \pi/b} e^{i \bfp \cdot {\bf x}} s(\bfp)
\ee
is coarse-grained spin at $\bfx$.
Actually, higher terms such as $O(\bar{s_b}^6)$ are present in $H(L,b)$
but they vanish as $L$ and $b$ become large.

Note that $< \sum_{\bfp} \bar{s_b}(\bfp)\bar{s_b}(-\bfp)>=\mbox{const}$
by definition and there are only two kinds of interesting terms.
If we set $b=L$, the term (\ref{hlbb}) is exactly the Binder's parameter
and we denote it by $B_L$.
The term (\ref{hlbc}) vanishes if we set $b=L$, so we must stop RG transform
at $b=L/2$. Then it becomes
\be
{
2\pi^2 <s( \bfk_1 )s(-\bfk_1)>
\over
2^{-D}< s(0)^2 +
2 s(\bfk_1 )s( -\bfk_1)>
},
\ee
where 
$\bfk_1\equiv(2\pi/L,0,\cdots,0)$.
For simplicity, we use the following quantity:
\be
C_L \equiv
{
 <s( \bfk_1 )s(-\bfk_1)> \over
 <s^2(0)> .
}
\label{cldef}.
\ee
Recently,
several different  parameters have been proposed as the second parameter
$C_L$
to estimate and eliminate the subleading scaling field.
We will review these works in Sec.II A.

Now consider the following matrix:
\be
{\partial (B_{bL},C_{bL} )\over
\partial (B_L,C_L)}
=
{
\partial (B_{bL},C_{bL} )\over
\barpa (\barbe_1,\barga_1)}\cdot
\left( {
\partial (B_L,C_L)
\over
\barpa (\barbe_1,\barga_1)}
\right)^{-1}
\nonumber \ee
\be
={
\partial ( B_{bL},C_{bL} )\over
\barpa (\barbe_b,\barga_b)}\cdot
{
\partial(\barbe_b,\barga_b)\over
\barpa (\barbe_1,\barga_1)
}
\cdot
\left( {
\partial ( B_L,C_L )
\over
\barpa (\barbe_1,\barga_1)}
\right)^{-1}
\ee
instead of
\be
{
\partial [\barbe(bL,bL),\barga(bL,bL/2)] \over
\partial [\barbe(L,L),\barga(L,L/2)]
}.
\ee

At the fixed point,
\be
{ \partial ( B_L,C_L )
\over
\barpa (\barbe_1,\barga_1) }
=
{ \partial ( B_{bL},C_{bL} )
\over
\barpa (\barbe_b,\barga_b) }
\ee
is satisfied and eigenvalues of
$\partial ( B_{bL},C_{bL} )/
\partial ( B_L,C_L )
$ coincide with that of
$\partial(\barbe_b,\barga_b)/
\barpa (\barbe_1,\barga_1)$ as long as
$
\partial ( B_L,C_L )
/\barpa (\barbe_1,\barga_1)
$ is a nonsingular matrix.

Thus  scaling behavior of renormalized parameters
can be extracted from  that of  $(B_L,C_L)$,  as long as
$|\partial ( B_L,C_L ) /\barpa (\barbe_1,\barga_1) |\neq 0$:
If we draw arrows from$(B_L,C_L)$ to $(B_{bL},C_{bL})$ 
in the $B_L$-$C_L$ plane,
the renormalization flow diagram of factor $b$ is obtained.
From this diagram, one can determine whether 
the MC result is asymptotic enough or not, 
by checking whether  $(B_L,C_L)$  converge
to a fixed point.
When a subleading scaling field is expected to be very large,
such as in the recent Monte Carlo studies of 
five-dimensional (5D) Ising model 
\cite{luijten,mon,parisi5d},
this RG flow diagram is very useful
compared to a many parameter fit to a single observable $B_L$.

A linearized RGT matrix
${\bf R }_b=\partial ( B_{bL},C_{bL} )/
\partial ( B_L,C_L )$
is calculated, for example, from linear fitting
\be
\left(
\begin{array}{c}  
B_{bL}(\alpha,\beta,\gamma) \\
C_{bL}(\alpha,\beta,\gamma)
\end{array} \right)=
{\bf R}_b 
\left(
\begin{array}{c}
B_{L}(\alpha,\beta,\gamma) \\
C_{L}(\alpha,\beta,\gamma)
\end{array} \right)
+ 
\left(
\begin{array}{c}
B_0 \\ C_0
\end{array} \right)
\label{lfit},\ee
where  ${\bf R} _b $ and $(B_0,C_0)$
are fitting parameters
and  we use  values of $(B_{L},C_{L})$ and $(B_{bL},C_{bL})$
at several different 
parameters $\alpha,\beta,\gamma$ near the fixed point.
Selection of this parameter is a delicate problem:
when  it is too close to the fixed point,
$(B_L,C_L)$ and $(B_{bL},C_{bL})$ become very close to each other 
and the RG flow is buried in statistical errors,
while when it is far from the fixed point,
nonlinear dependence of  $(B_L,C_L)$ on $(\alpha,\beta,\gamma)$
induces a systematic error. Thus the parameter range for the fitting
should be determined carefully.

Of course ${\bf R}_b$
can be calculated from
$\alpha,\beta,\gamma$ derivatives of $B_L$ and $C_L$
at the fixed point. However, derivatives with respect to 
irrelevant direction vanish as $L$ becomes large and buried in
statistical errors.
Thus values of $B_L,C_L$ at parameters well apart from
the fixed point along  an irrelevant direction are needed to calculate
the second eigenvalue.

\subsection{The second parameter in literature}
Here we compare our definition of $C_L$,
 ``the second parameter'' 
(the first one is the Binder's parameter),
  with preceding works.

Ballesteros {\it et al.} \cite{ballesteros}
used a finite-size correlation length defined below as
\be
\xi^2  \equiv  {
<\phi(0)^2>/<\phi(\bfk_1)\phi(-\bfk_1)> -1\over
\sin^2 (|\bfk_1|) },
\ee
which is related to $C_L$ as  $\xi^2 \sin^2(|\bfk_1|)= -1+1/C_L $.

Hasenbusch \cite{hasenbusch} used the ratio between the partition function of
a periodic and antiperiodic Hamiltonian.
In momentum representation,
the Hamiltonian for an antiperiodic boundary condition (APBC) is 
obtained by replacing  $\bfk$ summation of the lattice Hamiltonian
by $k_{\mu}=\pm \pi/L,\pm 3\pi/L,\cdots$ for each direction
$\mu$ to which the APBC is imposed.
Let us denote  the 
partition function of a periodic and antiperiodic Hamiltonian
by $Z_p$ and $Z_a$, respectively. It then reads:
\be
Z_a/Z_p= 
{ \int {\cal D}\phi \exp(-H_a) \over
\int {\cal D}\phi \exp(-H_p) }
\ee
\be=
{ \int {\cal D}\phi \exp(-H_p) \exp(H_p-H_a) \over
\int {\cal D}\phi \exp(-H_p) }
= < \exp(H_p-H_a)>_p
\ee
where $H_p$ and $H_a$ denote the 
periodic and antiperiodic Hamiltonian, respectively,
$<\cdots>_p$ denotes average with respect to $H_p$.
When the APBC is imposed to a direction of ${\bf e}_1$, it reads
\be
H_a-H_p \sim \sum_{\bfk}  
{2\pi |k_1| \over L} \phi(\bfk)\phi(-\bfk)
\ee
for large $L$.
One can see that $Z_a/Z_p$ has a similar form as that of $C_L$.
In the real-space RG scheme, both $C_L$ and $Z_a/Z_p$ can be
regarded as
an effective coupling  between two block spins
defined on $L\times L\times \cdots \times L/2$ block.

\section{ perturbation expansion at $D=4$}
Near the Gaussian fixed point,
finite-size behavior of 
$B_L$ and $C_L$ can be 
predicted from
finite-size perturbation theory  proposed by 
Chen and Dohm\cite{cd}.
Note that, when $D=4$, there is one kind of divergent 
subdiagram (Fig. \ref{fish}) whose factor is proportional to
$\left[\beta_1 (C_0+C_1 \ln L)/\gamma_1^2\right]$ at the critical region,
where  $C_0,C_1>0$ are some  constants.
Thus perturbation is restricted  to the range
$\beta_1 (C_0+C_1 \ln L)/\gamma_1^2\ll 1$ and 
one cannot set $L\rightarrow \infty$.
However, the  limit for
$L$  rapidly diverges as we approach the Gaussian fixed point,
and  good agreement between perturbation theory and
Monte Carlo data is expected for a certain parameter range and lattice size.
Then one can predict scaling behavior for large $L$,
which is far beyond the computational limit of Monte Carlo simulation,
from the perturbation theory.

Here we investigate finite-size behavior of 
$B_L$ and $C_L$ at the finite-size critical point (to one-loop order):
\be
\alpha= -12 { \beta \over \gamma} L^{-D}\sum_{\bfk \neq 0}
{1 \over 2 J(\bfk)}
\ee
where $J(\bfk)= \sum_\mu (1-\cos k_\mu ) $.

$B_L$ takes the so-called zero-mode value:
\be
 B_{\rm ZM}\equiv 
{\int d\Phi_0 \exp(-\Phi_0^4)\Phi_0^4 /\int d\Phi_0 \exp(-\Phi_0^4)
\over \left[
\int d\Phi_0 \exp(-\Phi_0^4)\Phi_0^2 /\int d\Phi_0 \exp(-\Phi_0^4) 
\right]^2
} \approx 2.1844. 
\ee

As for $C_L$, to one-loop order,

\be
C_L =  {\sqrt{\beta}\over 8\pi^2 \gamma}
\sqrt{ 1-36 {\beta\over \gamma^2} \sum_{\bfk \neq 0} [2J(\bfk)]^{-2} } .
\ee
For large $L$, $ \sum_{\bfk \neq 0} [2J(\bfk)]^{-2} \sim A_1 +A_2 \ln L$
where $A_1$ and $A_2$ are some positive constant.
Thus the plot of $(B_L,C_L)$ for a certain range of parameters including
critical temperature approaches $(B_{\rm ZM},0)$ as $L$ increases or 
$\beta$ decreases, indicating that the Gaussian fixed point is
infrared stable. However, approach to the point $(B_{\rm ZM},0)$ 
for increasing $L$ is extremely slow, 
and one cannot expect asymptotic Gaussian behavior 
in MC simulations, even when $L$ is very large.
Critical exponents estimated  from MC results may also differ
from the Gaussian (classical) value and depend on the bare parameters.
Thus one can extract only
very restricted information from MC simulations at the upper critical
dimension. 

\section{ Detail of Monte Carlo simulation}
We investigated the 
$L\times L\times L$ system with $L=8,16,32$ for $D=3$ and
the $L\times L\times L \times L$ system with $L=4,8,16$ for $D=4$,
imposing a periodic boundary condition on each directions.

We used the following Hamiltonian:
\be
H={\gamma \over 2}\sum_{<ij>} (s_i-s_j)^2 + {\alpha \over 2}  \sum_i s_i^2 
+\beta \sum_i s_i^4
\ee
where
$-\infty <s_i<\infty $ denotes the spin on the site $i$.
Since we cannot use the regularization condition $<s_i^2>=1$
before the simulation, we used the following one
$$
\frac{
\int d\phi \exp( -{\alpha \over 2} \phi^2 -\beta \phi^4) \phi^2
}{
\int d\phi \exp( -{\alpha \over 2} \phi^2 -\beta \phi^4)}=1 .
$$
We used several fixed $\alpha,\beta$ 
and tuned  $\gamma$ to reach the critical region.
Actual values of parameters are listed in the later sections.

For parameters well apart from the Gaussian fixed point,
$4 L^{D-2}$ single cluster flips \cite{wolff} and $16$ Metropolis sweeps
are performed between successive observations.
In the cluster-update stage, length of $s_i$ is kept fixed and
only its sign is changed.
In the Metropolis update step,
a new value for $s_i$ is chosen uniformly from a range
$\exp( -\alpha s^2 -\beta s^4)\geq \exp(-3.0)$.
By the cluster update, all spins are flipped four times on average between
observations.
For all observed quantities,
the correlation coefficient between  successively observed values
was less than $0.2$.

As we approach the Gaussian fixed point, parameters 
behave as follows:
\be
\alpha \sim 0 \comm \beta \sim \mbox{const} \comm \gamma \rightarrow \infty.
\ee
Thus the nearest neighbor coupling $\gamma s_i s_j/2$
tends to diverge. Since the bond-cutting probability of
Wolff's algorithm is $\exp( -\gamma s_i s_j)$,
the cluster update tends to
end up with flipping the whole system and does not accelerate the simulation.
In these cases we increased the number of Metropolis sweeps until the above
mentioned condition is satisfied.

Thermal averages of observables  at  $\gamma$
slightly away from the actually used value were calculated
using reweighting techniques \cite{hist}.
For all system sizes, at least $ 0.8\times 10^5$ observations were done
after thermalization, and 
statistical errors were estimated by the jack-knife procedure.
Multiplicative lagged Fibonacci sequence 
$R_t=R_{t-9689}\times R_{t-4187}  \,\,\, ( {\rm mod}  2^{31})$
was used as a random number generator.
All runs were  performed on  VPP-300 at JAERI.

\section{ Result for $D=3$}

Simulations were performed at the following parameters 
(see also Fig. \ref{ab0}) for $L=8,16,32$;
\begin{description}
\item[Ising case:] $\phi(\bfx)=\pm 1 \comm \gamma=0.2217;$
\item[Case A:] $\alpha/2=-2.6159\comm \beta=1.0948 \comm\gamma=0.3040;$
\item[Case B:] $\alpha/2=-1.6655\comm \beta=0.6935 \comm\gamma=0.3545;$ 
\item[Case C:] $\alpha/2=-0.8786\comm \beta=0.3938 \comm\gamma=0.4700;$
\item[Case D:] $\alpha/2=-0.5209\comm \beta=0.2713 \comm\gamma=0.6260$ .
\end{description}

Figure \ref{rg3mc} 
shows the RG flow diagram of $B_L$ and $C_L$ near the Wilson-Fisher 
RG fixed point.
All lines are drawn from
$(B_L,C_L)$ to $(B_{2L},C_{2L})$.
Dashed and solid lines correspond to   
$L=8$ and $L=16$, respectively.
One can see that 
the RG flow shown in Fig. \ref{rgf3} is well
reproduced.
Moreover,
the linear fitting procedure (\ref{lfit})
provides an estimate for
critical exponents  as
$\nu=0.69(3)$, $\omega=0.74(10) $ from $L=8,16$ data and
$\nu=0.653(10), \omega=0.7(2)$ from $L=16,32$ data,
which is in agreement with the most recent 
$\epsilon$-expansion result $\nu=0.6305(25),\omega=0.814(10)$ \cite{zinn}
and the Monte Carlo result $\nu=0.6296(3),\omega=0.845(10)$ \cite{hasenbusch}.
Thus one can see that the two observables $B_L$ and $C_L$ are
enough to capture the essential RG flow.

\section{ Result for $D=4$}
\begin{figure}
\centerline{
\epsfysize=5cm
(a) \epsfbox{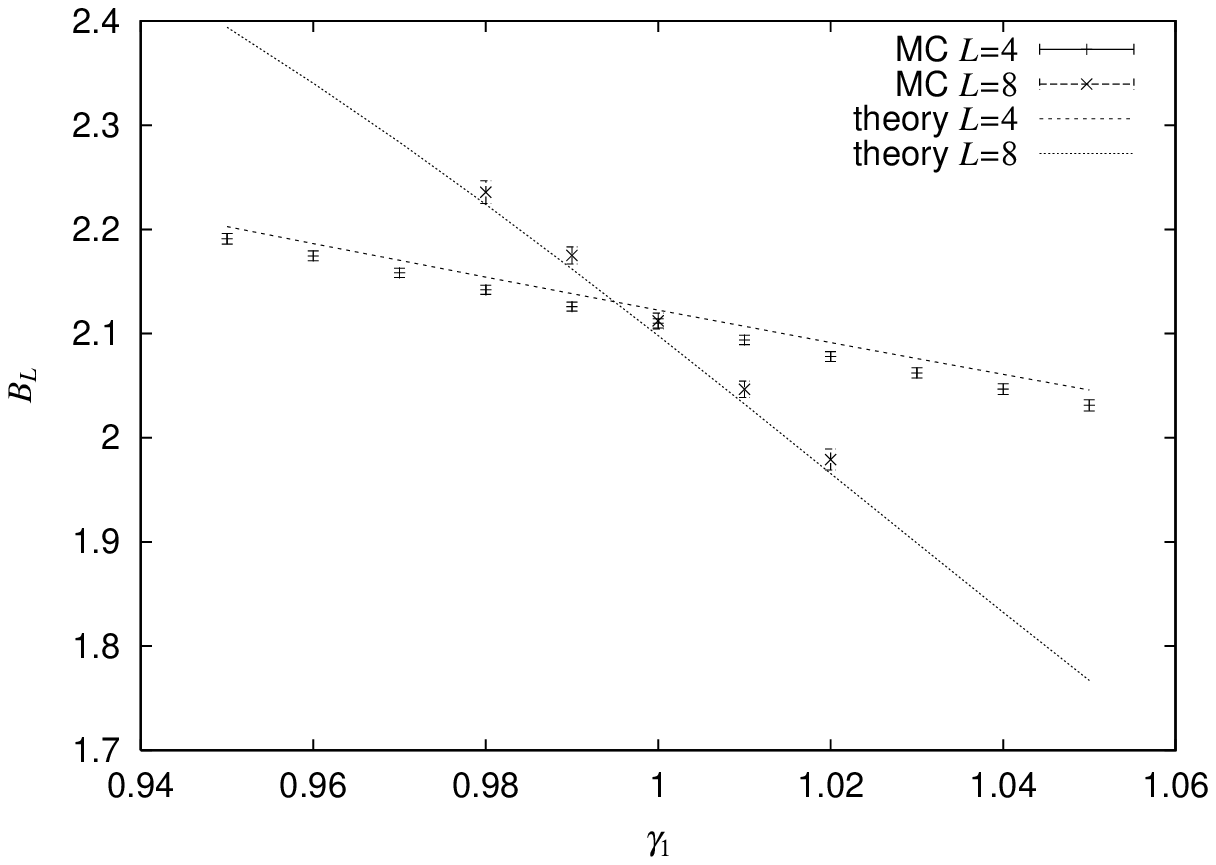}
\epsfysize=5cm
(b) \epsfbox{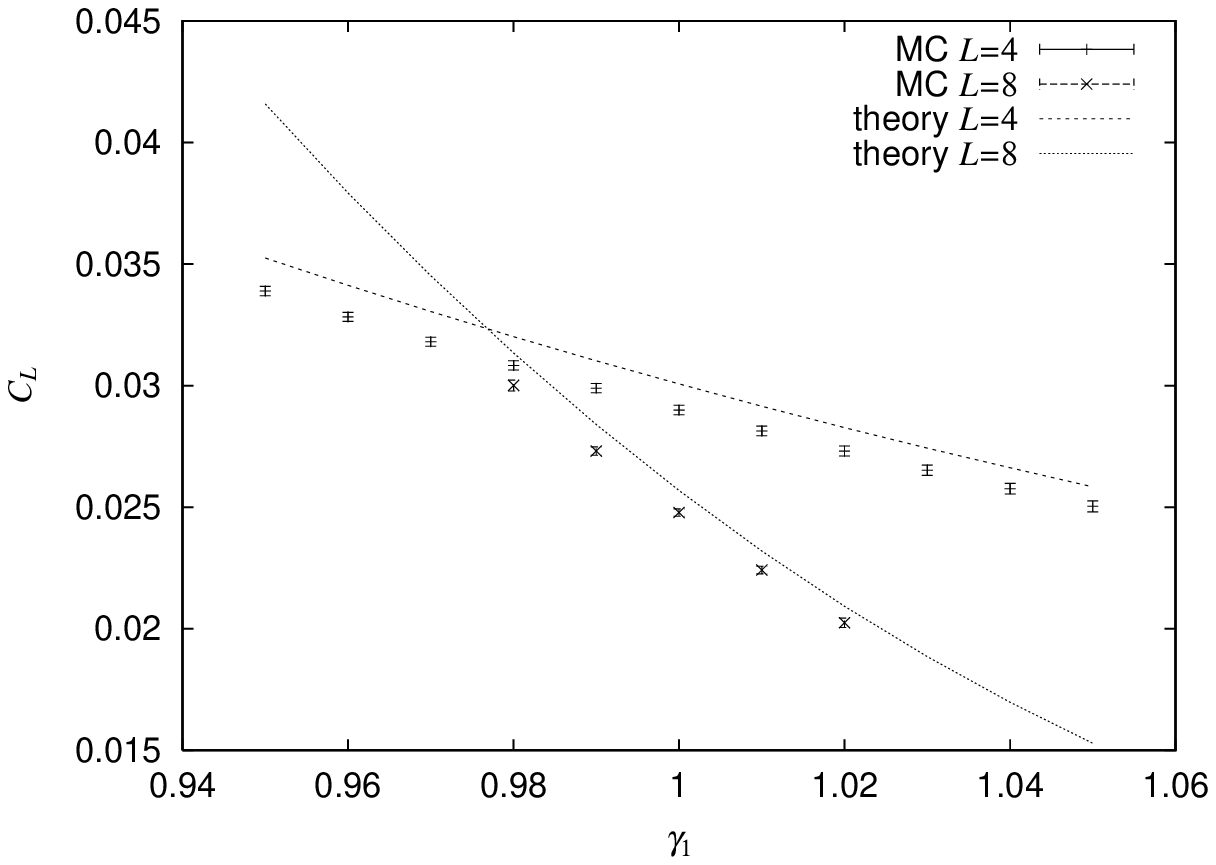}
}
\caption{
Plot of (a) $B_L$ and (b) $C_L$ against $\gamma_1$ for the Case B
in $D=4$.
``MC'' denotes Monte Carlo result and ``theory'' denotes
finite-size perturbation result.}
\label{bl4d}
\end{figure}

Simulations were performed at
the following parameters,
\begin{description}
\item[Ising case:] $\phi(\bfx)=\pm 1 \comm \gamma_1=0.1495;$
\item[Case A:] $\alpha/2=-0.3226\comm \beta=0.2082 \comm\gamma=0.54;$
\item[Case B:] $\alpha/2=-0.1383\comm \beta=0.1530 \comm\gamma=1.0$ 
\end{description}
for $L=4,8,$ and $16$.
Near the Gaussian fixed point (Case B),
there is a severe, critical slowing down (owing to large fluctuation of
the $\bfk=0$ mode), which cannot be removed
by the cluster update, and we did not perform $L=16$ simulation.
For Case B, 
perturbation expansion agrees well with the MC result:
Fig.\ref{bl4d}(a) and \ref{bl4d}(b) 
show the plot of $B_L$ and $C_L$, respectively,  against $\gamma_1$ for 
fixed $(\alpha,\beta)$, together with perturbation results.
Note that there are no free parameters to
be fitted,  unlike the Ising case \cite{5dcdmc}, and 
the agreement is both qualitative and quantitative.

Figure \ref{rgmc4} shows a RG flow diagram of $B_L$ and $C_L$ 
obtained from  MC simulations.
All lines are drawn from
$(B_L,C_L)$ to $(B_{2L},C_{2L})$.
Dashed and solid lines correspond to   
$L=4$ and $L=8$, respectively.
Simulations at a parameter
closer to the Gaussian fixed point than Case B 
are  very difficult owing to aforementioned critical slowing down.
Instead, finite-size perturbation 
provides reliable results near the Gaussian  fixed point
and it indicates that
the plot of $(B_L,C_L)$ approaches the Gaussian  fixed point
as $L$ increases.
Thus one  can conclude  that
there is no RG fixed point except for the infrared-stable 
Gaussian fixed point.

\section{conclusion}
The renormalization-group flow diagram
obtained by the method presented in this paper
provides qualitative information such as
the stability of a specific RG fixed point against some perturbation,
as well as 
quantitative improvement of the 
estimated value of a critical exponent
by eliminating leading correction-to-scaling terms.
However, in lattice models, there exists $O(1/L)$ systematic error
owing to the substitution of integral by finite summation of $1/L$ mesh,
and
one cannot get rid of this as long as the
finite lattice system is concerned.


Our method can  be easily
extended  to  $\phi^4$ models with several or unusual 
quartic coupling constant(s):
\be
H=\int dx (\nabla \phi)^2 +\alpha \phi^2 +\sum_n
\beta_n \sum_{ijkl}C_{ijkl}(n) \phi_i \phi_j \phi_k \phi_l
,\ee
such as the
chiral $O(2n)$ model of a
 triangular antiferromagnet \cite{kawamura}, 
and the Ginzburg-Landau model of a
type-II superconductor under  
a weak or strong magnetic field, with or without
point/columnar impurities (see \cite{brezin,hu} for
transition of a pure system under a strong field).
The multiple quartic term tends to generate
an irrelevant operator whose correction exponent is very small,
making it difficult to observe asymptotic behavior in MC simulations.
Thus critical behavior of these models has been a
controversial issue and 
application of the MCRG method  
to these models seems very interesting.
A RG flow diagram of regularized quartic terms 
$ <C_{ijkl}(n) \phi_i \phi_j \phi_k \phi_l>/<\sum_i \phi_i^2>^2 $
will reveal the critical behavior, as in Ref.\cite{q4potts} of 
the $Q=4$
antiferromagnetic Potts model.

Another important problem is the estimation  of the
correction exponent
for the $\phi^6$ term at the WF fixed point.
An $\epsilon$ expansion analysis indicates that it becomes
 positive at the 
WF fixed point \cite{phi6}.
However, whether the exponent 
is larger or smaller than $\omega \approx 0.8$ 
of $\phi^4$ theory, 
which we assumed as a {\it leading} correction,  
should be confirmed numerically.
Similarly, the effect of sixfold anisotropy 
on the critical behavior of a three-dimensional XY model is another
interesting subject since  the anisotropy  
is expressed by the $O(\phi^6)$ term.

\myfig{rgf3}{5cm}
{Renormalization flow of regularized parameter in $D=3$.}

\myfig{rgf4}{5cm}
{Renormalization flow of regularized parameter in $D=4$.}

\myfig{fish}{5cm}
{Divergent subdiagram in $D=4$.}

\myfig{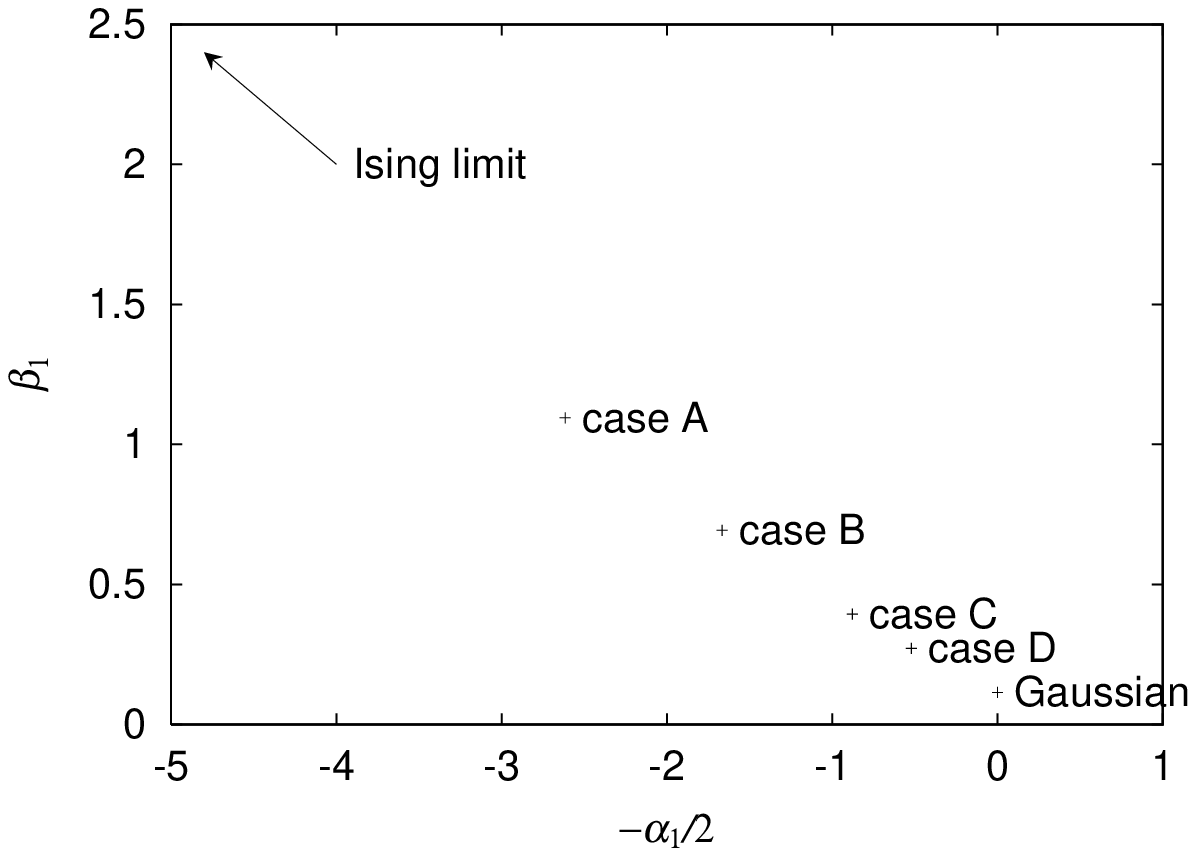}{5cm}
{Parameters at which simulations were performed.}
\myfig{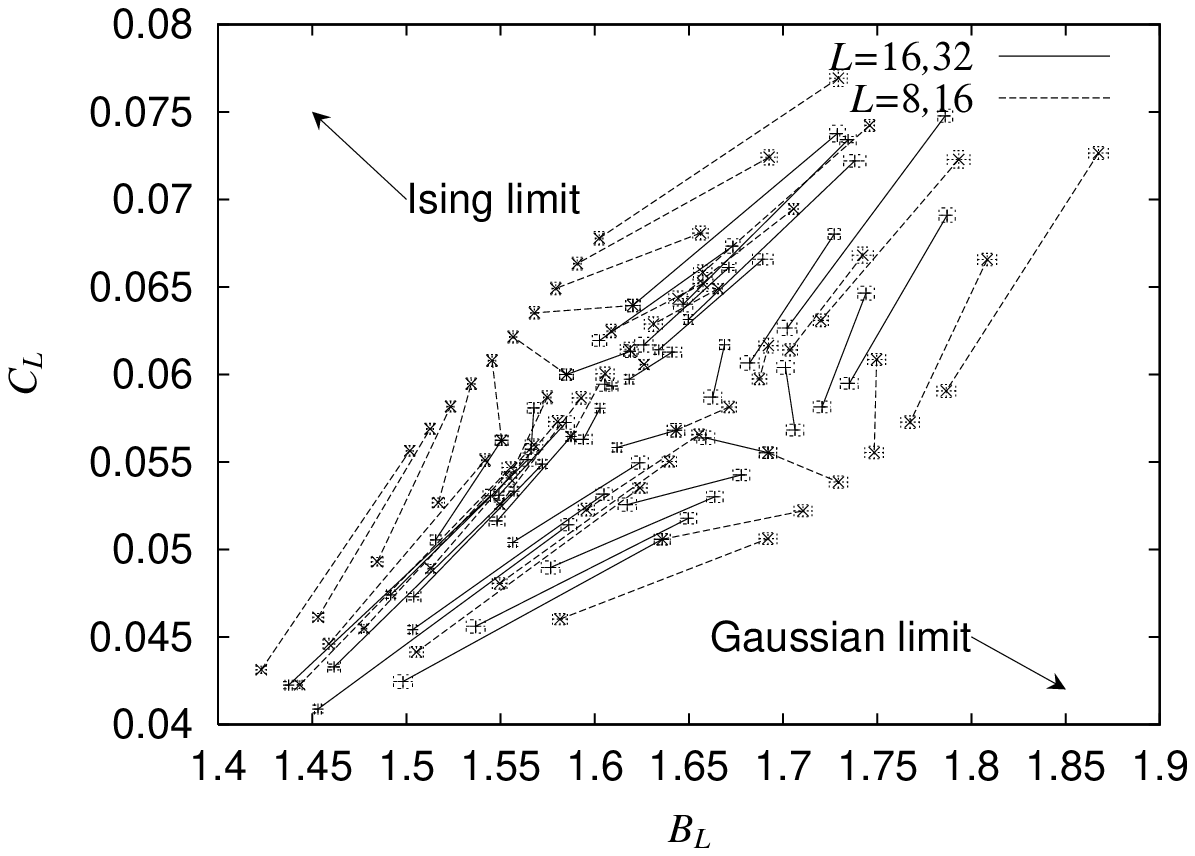}{6cm}
{RG flow near the Wilson-Fisher fixed point in  $D=3$. 
All lines are drawn from
$(B_L,C_L)$ to $(B_{2L},C_{2L})$.
}

\myfig{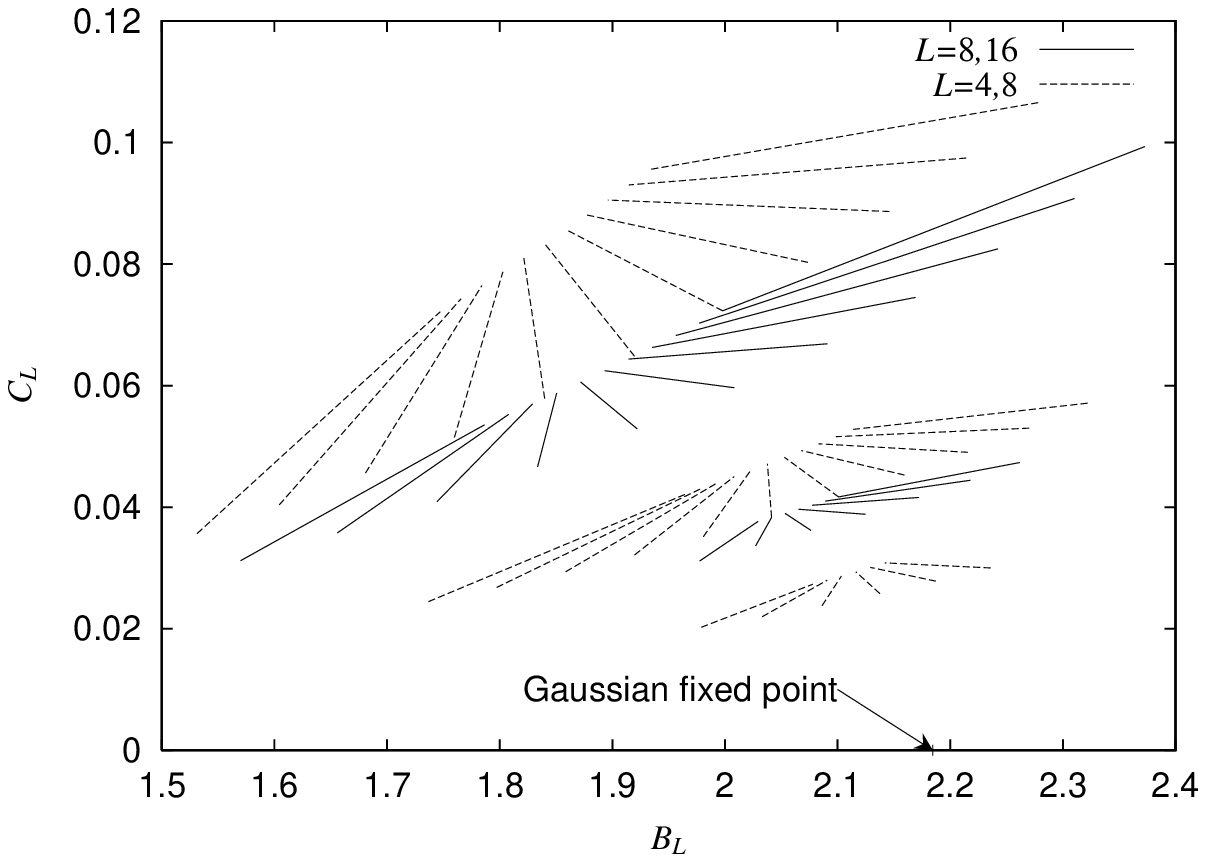}{6cm}
{RG flow of $B_L$ and $C_L$ , obtained by MC 
simulation in  $D=4$. 
All lines are drawn from $(B_L,C_L)$ to $(B_{2L},C_{2L})$.
Dashed and solid lines correspond to   
$L=4$ and $L=8$, respectively.
}

\end{document}